\documentclass[12pt,a4paper]{article}

\usepackage{graphicx}
\usepackage{algorithm}
\usepackage[noend]{algpseudocode}
\usepackage{color}
\usepackage{amssymb, amsmath, amsthm, amsfonts} 

\title{Optimal policy design for the sugar tax}
\author{K.~Geyskens \and A.~Grigoriev \and N.~Holtrop\\ 
\small{Maastricht University School of Business and Economics}\\ 
\small{P.O.Box 616, 6200 MD Maastricht, The Netherlands}\\ \small{\{K.Geyskens;A.Grigoriev;N.Holtrop\}@maastrichtuniversity.nl} 
\and A.~Nedelko\\ 
\small{Higher School of Economics}\\ 
\small{P.O.Box 33, Kirpichnaya street, Moscow, Russia}\\  
\small{anedelko@hse.ru}
}
\date{}
\begin{document}
\maketitle

\begin{abstract}
Healthy nutrition promotions and regulations have long been regarded as a tool for increasing social welfare. One of the avenues taken in the past decade is sugar consumption regulation by introducing a sugar tax. Such a tax increases the price of extensive sugar containment in products such as soft drinks. In this article we consider a typical problem of optimal regulatory policy design, where the task is to determine the sugar tax rate maximizing the social welfare. We model the problem as a sequential game represented by the three-level mathematical program. On the upper level, the government decides upon the tax rate. On the middle level, producers decide on the product pricing. On the lower level, consumers decide upon their preferences towards the products. While the general problem is computationally intractable, the problem with a few product types is polynomially solvable, even for an arbitrary number of heterogeneous consumers. This paper presents a simple, intuitive and easily implementable framework for computing optimal sugar tax in a market with a few products types. This resembles the reality as the soft drinks, for instance, are typically categorized in either regular or no-sugar drinks, e.g. Coca-Cola and Coca-Cola Zero. We illustrate the algorithm using an example based on the real data and draw conclusions for a specific local market.
\end{abstract}


\section{Introduction}\label{intro}
Since the 2000s, interest in the proper nutrition promotion has dramatically increased in many countries. According to the World Health Organization (WHO), worldwide obesity has nearly tripled since 1975. In 2016, more than 1.9 billion adults were overweight, with over 650 million of these adults being obese. Moreover, 41 million children under the age of 5 were overweight or obese in 2016~\cite{who}. In addition to obesity, adults, children and adolescents often face risks such as depletion, stunted growth, lack of vitamins and minerals, nutritional non-communicable diseases (heart disease, stroke, diabetes, certain cancers) that affect their health in the short and long term periods~\cite{who}. Therefore, the formation of a proper nutrition culture is very important, especially in youth, since all the human habits and life values are formed at a young age \cite{ferreira2007environmental}. During adolescence, young people adjust their lifestyle easier than adults. Thereby, nutrition habit is one of the basic elements for their health in future.

Solving obesity problems is also a challenging task for the governments, as inadequate nutrition increases health care costs, reduces productivity and slows economic growth. These consequences, in turn, are the basis for permanent poverty and poor population health \cite{who}.

In contrast, the main concern of the companies is the average market share and/or profit~\cite{smith2013managing}. This incentives are not always directed at healthy products, and it is always a challenge for a government to introduce and to maintain mechanisms stimulating the companies to promote proper nutrition. Clearly, there might exist opposing interests, when the companies focus on purely financial indicators, while the government seeks to improve the social welfare. In this interaction, the end-consumers play crucial role. On the one hand, consumers choose products guided by various marketing stimuli such as advertising, pricing, and branding \cite{solomon2012consumer}. This way consumers strongly support the companies. On the other hand, the governmental and societal information provision programs create awareness of proper nutrition importance, shift the demand towards healthy products and, as a result, customer valuations for healthy products become higher than for unhealthy ones. This way consumers strongly support the government. Therefore, a combination of information provision and direct regulations, e.g. extra taxation for unhealthy products, is a powerful instrument of the government to improve the social welfare.

\subsection{Nutrition promotion instruments}\label{NP analysis}
In this section we describe possible nutrition promotion mechanisms for a government and for a company/firm. Later, some of the mechanisms, namely, taxes and prices, will be explicitly introduced as variables in the utility functions of the companies and consumers, respectively. 

The government is typically the first mover which sets monetary and/or non-monetary product/market regulations. Among monetary regulations, the most popular ones are \emph{taxes} (for unhealthy products), \emph{subsidies} (for healthy products), and \emph{caps} (maximal price for a product). These regulations directly influence the utility function of a company and rarely affect the consumer utility functions. Furthermore, government can use non-monetary instruments to stimulate nutrition consumption, e.g., certification, labeling, obligatory description of ingredients and nutritional value on the packages~\cite{stevenson2012environmental,minkov2015type}. More specifically, the government can impose an obligation on firms to use particular size, color and shape for nutritious food packages or for price tags. For instance, recent research shows that the consumers perceive products in vivid packaging as less healthy than food in muted color packages \cite{mead2018package}. \emph {Minimal font size} of ingredients' inscription and nutritional value also can be used by government as non-monetary regulation. This tool may attract consumers' attention on containment of harmful to health ingredients such as sugar, preservatives and dyes. The last but not the least tool is \emph {point-of-sales merchandising}. One study showed that joint presentation of healthy and unhealthy products forces consumers to choose the first one because such way increases guilt and the difficulty of social justification \cite{okada2005justification}. Consequently, merchandising can be an effective instrument which may force people to buy healthy products. These non-monetary regulatory mechanisms serve for information provision and do have influence on the consumer utility functions, though the effect of the information provision is sometimes is not immediate. The company utility functions are rarely directly affected by the non-monetary governmental regulations.

In turn, a company/firm (producer or retailer) is the follower. Knowing the governmental regulations, companies maximize the profit applying their toolkit to influence the consumer behavior. It should be noted that interests of producers and retailers can be different, but within the framework of this research we do not distinguish these two players. It is widely accepted that the most powerful tool of a firm is the \emph{price} set for a product. Next to the price, \emph{discounts} is the most popular instrument. To illustrate the possible efficiency of this tool, prior research has provided results of experiments where price reduction can be a reason of increased consumption~\cite{geliebter2013comparison,ball2015influence}. For example, \cite{ball2015influence} have found that a 20\% discount for fruit and vegetables categories caused increased purchasing per household of 35\% for fruit and 15\% for vegetables. Furthermore, it is proved that even temporary discounts can stimulate proper nutrition consumption. In one study~\cite{geliebter2013comparison}, there were three periods (baseline - no discount, intervention - 50\% discount, follow-up - no discount) during which obese respondents were buying fruits and vegetables during this periods. As a result, purchasing of fruits and vegetables during follow-up period became higher than during baseline. This research demonstrates effectiveness of applying discounts to increase proper nutrition consumption. At the same time, firms may use such non-monetary instruments as \emph{availability in stores}, \emph{merchandising}, and \emph{package design}~\cite{glanz2012retail}. Availability in stores means that wide access to proper nutrition food increase nutrition consumption. It happens because people prefer to buy what is in every supermarket instead of trying to find something special for everyday meals~\cite{desai2003consumer,morales2005giving}. Merchandising was already discussed in the context of the governmental tools and it can also be effective on the company's level. Moreover, unusual packages can be used by marketers to increase demand for proper nutrition~\cite{rettie2000verbal}.

\subsection{Problem statement and basic assumptions}
In this study we assume that (1) effective information provision programs take place; (2) consumers are aware of proper nutrition importance; and (3) consumers already formed their utility functions (product valuations) and in the near future they are not going to change their preferences significantly. Notice, without these assumptions the efficiency of direct regulations is questionable. We take the consumer utility functions as granted. This is also a valid assumption, given a number and variety of consumer behavior models available in the literature. In the end of the paper, we provide an insightful example and analysis based on a specific consumer behavior model and actual purchase data. Such data can be routinely obtained from the market research agencies such as Nielsen, GfK and Kantar. We base our example and experiments on the data provided by the latter party.

Specifically targeting the optimal regulations of the soft drinks markets, this research focuses on the most popular, powerful and long-run instruments of the government and of the firm, namely on the sugar tax and prices, respectively. In the past decade, many countries initiated sugar consumption regulation by introducing a sugar tax. Such a tax increases the price of extensive sugar containment in products such as soft drinks. In this article we consider a typical problem of optimal regulatory policy design, where the task is to determine the sugar tax rate maximizing the social welfare. We model the problem as a sequential game represented by the three-level mathematical program. On the upper level, given the utility functions of the soft drink producer/retailer and consumers, the government decides upon the sugar tax rate with the goal to optimize the social welfare. On the middle level, given the sugar tax rate and the utility functions of the consumers, a company decides on the product prices maximizing the company utility. On the lower level, given the product prices, consumers decide upon their preferences towards either sugar containing or sugar-free products. All utility functions and the social welfare are taken from the classic economic literature. This ensures the generality of the approach and applicability to not only the soft drinks markets and sugar taxes, but rather to a broad variety of the markets needing governmental regulations. 

\subsection{Contemporary sugar tax practices}
The sugar tax means each liter of sugary drink will have an extra tax charge up to 50\%, depending on how much sugar is in the drink\footnote{https://www.bbc.com/news/health-35824071}. Tax rates depend on government policy in a country and can be expressed in percentage or in monetary units. Moreover, taxation schemes vary from country to country, see Table~\ref{tab:countries}. Nowadays, there are two common schemes: a one-level tax rate and a multi-level tax rate. In the case of a one-level tax rate, government establishes a single tax rate for all drinks containing sugar, or for drinks with a sugar containment above a specific threshold. Such approaches are used in France, Chile, Mexico, Belgium, Colombia, India, Portugal, Saudi Arabia, UAE, USA, South Africa. Alternatively, several countries (e.g. Thailand, Ireland, UK) apply a multi-level taxation scheme, which assumes different tax rates according to sugar content\footnote{https://www.beveragedaily.com/Article/2017/12/20/Sugar-taxes-The-global-picture-in-2017}, see Table~\ref{tab:countries}.

\begin{table}[h!]
\footnotesize
\centering
\begin{tabular}{l|l|l} 
  \hline
  \bf{Country} & \bf{Tax rate} & \bf{Effective}\\
  & & \bf{since}\\  
  \hline
  France & 7.53 euro per 100 liters & 2013   \\\hline
  Chile & $>6,25$ grams of sugar per 0.1 liter $\rightarrow$ 16\% & 2014\\
  \hline
  Mexico & 1 peso per liter & 2014\\
  \hline
  Belgium & 3.7284 euro per 100 liters & 2013\\
  \hline
  Colombia & 20\% per liter & 2016\\
  \hline
  India & 40\% & 2017\\
  \hline
  Portugal & $>80$ grams of sugar per liter $\rightarrow$ 16\% & 2017\\
  \hline
  Thailand & 14\% + 5-stage sugar tax according to & 2017\\
  & sugar content\\ 
  \hline
  Saudi Arabia & 50\%  & 2017\\
  \hline
  UAE & 50\%  & 2017\\
  \hline
  USA (several cities)  & 1–2 cents per ounce  & 2017\\
  \hline
  Ireland & 5–8 grams of sugar per 0.1 liter $\rightarrow$ 21 cents;\\   
  & $>8$ grams of sugar per 0.1 liter $\rightarrow$ 31 cents  & 2018\\
  \hline
  South Africa & $>4$ grams of sugar per 0.1 liter $\rightarrow$ 2.1 cents\\ 
  & per gram of sugar per 0.1 liter  & 2018\\
  \hline
  UK & 5–8 grams of sugar per 0.1 liter $\rightarrow$ 18\%;\\  
  & $>8$ grams of sugar per 0.1 liter$ \rightarrow$ 24\%  & 2018\\
  \hline
\end{tabular}
\caption{\label{tab:countries} Sugar tax rates across countries}
\end{table}

In this paper we assume a one-level tax rate for any positive sugar containment in the drink. This is the current sugar tax practice in many countries, e.g., France, Mexico, Belgium, Colombia, India, Saudi Arabia, UAE and USA.The approach is straightforwardly extendable to a multi-level tax rate with a constant number of (a few) levels. 

\section{Definitions and mathematical model}
Since the tax per cent varies widely from country to country, it makes sense to develop a general mathematical model determining the taxation mechanism maximizing the social welfare. Such a model should coordinate the interests across the three players: government, firms and consumers. As a starting point for the model the utility functions of the players are defined. The government utility is usually understood as \emph{social welfare}, see~\cite{bernoulli2011exposition}. Let the social welfare be referred as $W$. It is expressed as the total utility of consumers and firms plus the tax:
\begin{displaymath}
W=U_{c} + U_{f} + T,
\end{displaymath}
where $U_{c}$ is the total utility of all consumers, $U_{f}$ is the total utility of all firms, and $T$ is the total tax collected.

The total utility of the consumers $U_c$ is defined as follows. Let the set of consumers be denoted by $N$. For simplicity of presentation, consider a market with only two products: one is containing sugar and another one is sugar-free. Let the product containing sugar be indexed with 1, and the sugar-free product be indexed with 0, and let $M=\{0,1\}$ be the product index set. Later we explain how to generalize the model and how to adjust the algorithms in the case of several (a constant number of) products. Let $u_{i,j}$ be an individual utility of consumer $i\in N$ for product $j\in M$, and let $x_{i,j}$ be a binary decision variable taking value 1 if consumer $i$ prefers product $j$ to any other products and 0 otherwise. Assuming consumer's rationality, we have $x_{i,k}=1,\ i\in N, k\in M$, if only if $u_{i,k}=\max_{j\in M}u_{i,j}$ and $u_{i,k}\geq 0$. Here, for all consumers $i\in N$ we assume that $\sum_{j\in M}x_{i,j}\leq 1$. For this assumption to be true,  the ties on the maximal utilities are broken in favor of the company revenue --- this is also a folklore economic assumption. Clearly, if a consumer has negative utilities for all products, she does not purchase anything, i.e., $\sum_{j\in M}x_{i,j}=0$.

In the literature on marketing, there are various models describing behavior and individual utilities of the consumers for a product. Vast majority of the models are linear in the product price. For instance, consumer utility function suggested in~\cite{healthclaims2017} consists of (1) a constant which includes different psychological, economical, and sociological factors such as brand loyalty, consumer's budget, readiness to pay etc; (2) a term depending on the number of claims and nutritional values which are written on product packages, e.g., labels as "low in fat", "high in fiber" and other which may influence consumer's choice; and (3) deduction of the price related factor: 
\begin{displaymath}
u_{i,j} = \left(\beta_{i,j}+\beta_{1}\cdot \overline{NrClaims}+\beta_{2}\cdot \overline{NutrVal} - \beta_{3}\cdot p_j\right)^+.\
\end{displaymath}
Here, $\beta_{i,j},\ \beta_{1},\ \beta_{2},\ \beta_{3}$ are the coefficients determined by a multinominal choice model\footnote{https://eml.berkeley.edu/books/choice2.html}, $\overline{NrClaims}$ is the average number of claims on product packages in category, $\overline{NutrVal}$ is the average nutritional value in category, and $p_k$ is the price of product $j$. Notice, the intercept $\beta_{i,j}$ is a constant including factors causing heterogeneity across the consumers. Finally, $$U_c=\sum_{i\in N}\sum_{j\in M} \ln (1+u_{i,j})\cdot x_{i,j},$$ under a classic assumption of diminishing marginal utilities of consumers, see, e.g.,~\cite{bernoulli2011exposition}.

The consumer behavior model in~\cite{healthclaims2017} does not assume price elasticity of the demand. This is a reasonable assumption for the present research as for the soft drinks, the volume of purchase depends rather on the preferences and demographic characteristics of the consumer's household than on the product price. Therefore, we may assume that the demand of a consumer $i\in N$ is given in two quantities: if the consumer prefers sugar-free drink, the realized demand is $D_{i,0}$, and if the sugary drink is preferred, the demand is $D_{i,1}$. Then, the utility of a company $U_f$ is defined by its revenue $$U_f=\left(\sum_{i\in N}\sum_{j\in M}\ D_{i,j}\cdot p_j\cdot x_{i,j}\right)\ -\ T,$$ while the tax $T$ is defined by $$T=\alpha\cdot \sum_{i\in N}\ D_{i,1}\cdot p_1\cdot x_{i,1},$$ where $0\leq \alpha \leq 1$ is the sugar tax rate established by the government.

Thus, the entire mathematical model is represented by the three-level program:
\begin{equation}
    \max_{0\leq \alpha\leq 1}\ \sum_{i\in N}\sum_{j\in M}\ \left(\ln (1+u_{i,j})\ +\ D_{i,j}\cdot p^*_j\right)\cdot x^*_{i,j}
\end{equation}
subject to
\begin{equation}
    p^*\ =\ \arg\max_{p\geq 0}\ \left(\sum_{i\in N}\ D_{i,0}\cdot p_0\cdot x^*_{i,0}\right)\ +\ (1-\alpha)\cdot \left(\sum_{i\in N}\ D_{i,1}\cdot p_1\cdot x^*_{i,1}\right)
\end{equation}
subject to
\begin{equation}
    x^*\ =\ \arg\max_{x}\ \sum_{i\in N}\sum_{j\in M}\ u_{i,j}\cdot x_{i,j}
\end{equation}
subject to
\begin{equation}
    u_{i,j} = \left(\beta_{i,j}+\beta_{1}\cdot \overline{NrClaims}+\beta_{2}\cdot \overline{NutrVal} - \beta_{3}\cdot p_j\right)^+,\
\end{equation}
\begin{equation}
    \sum_{j\in M} x_{i,j}\leq 1 \ \ \ \ \ \forall i\in N,
\end{equation}
\begin{equation}
    x_{i,j}\in \{0,1\} \ \ \ \ \ \forall i\in N, j\in M.
\end{equation}
On the first level, the government decides upon the sugar tax rate $0\leq \alpha\leq 1$ to maximize the social welfare. On the second level, given the tax rate $\alpha$, the company decides on the prices $p_0\geq 0$ and $p_1\geq 0$ to maximize the company's revenue. On the third level, given the prices $p$, the consumers select preferred products maximizing their utility. 

\section{Solution of the three-level program}
In this section we provide intuition and sketch the algorithms solving the three-level mathematical problem (1)-(6). Then, we present the formal pseudo-codes of the algorithms. As the algorithms are rather straightforward and very intuitive, we leave it for a reader to prove their correctness. 

To find the optimal tax rate, we first suggest to find all potentially optimal pricing strategies of the company. Notice, the consumer utility functions are linear in prices. Therefore, for every consumer $i\in N$, the \emph{preference half-spaces} are determined by inequalities $\beta_{i,0}-\beta_3 p_0\geq \beta_{i,1}-\beta_3 p_1,$ where the consumer prefers the sugar-free drink, and $\beta_{i,0}-\beta_3 p_0\leq \beta_{i,1}-\beta_3 p_1,$ where the consumer might prefer the sugary drink. The \emph{indifference hyperplane} $\beta_{i,0}-\beta_3 p_0= \beta_{i,1}-\beta_3 p_1$ does, actually, represent the prices where the consumer is indifferent which product to purchase. Next to indifference hyperplanes, let us introduce the \emph{budget hyperplanes}: $\beta_{i,0}=\beta_3p_0$ and $\beta_{i,1}=\beta_3p_1$, as the consumer purchases a product only if her utility for the product is non-negative, i.e., $\beta_{i,0}\geq\beta_3p_0$ or $\beta_{i,1}\geq\beta_3p_1$. The union of all indifference and budget hyperplanes splits the price space in many polyhedras. Let us refer to a polyhedra as a \emph{choice region} if the interior of the polyhedra is not intersected by any of the indifference or budget hyperplanes. 

By construction, for all consumers, their preferences for a product remain the same for any prices taken from the same choice region. Then, by linearity of the revenue function in prices, only vertices of the choice regions might become optimal pricing strategies of the company. Therefore, having $O(n)$ indifference and budget hyperplanes for only two products, where $n$ is the number of consumers, we may have at most $O(n^2)$ potentially optimal pricing strategies. When the number of products increases to $m$, the number of potentially optimal strategies increases to $O(n^m)$. Thus, keeping $m$ a small constant, brute-force enumeration of potentially optimal pricing strategies remains polynomial and can be efficiently implemented. Moreover, since the consumer utilities are independent on the sugar tax, for any choice of the tax rate, one of the computed pricing strategies will be optimal. Hence, the first step towards the optimal policy design is computing all potentially optimal prices, which is done by Algorithm 1 below.

\begin{algorithm} [h!]
\footnotesize
\caption{Computing potentially optimal prices}\label{alg:optprices}
\begin{algorithmic}[1]
\State {\bf Input}: $\overline{NrClaims}$, $\overline{NutrVal}$, $\beta_1$, $\beta_2$, $\beta_3$, $\beta_{i,j}$ for all $i\in N,\ j\in \{0,1\}$ 
\State $\mathcal{P}:=\emptyset$ \Comment Set of potentially optimal prices
\State Create a list $\mathcal{B}$ of budget hyperplanes: $\beta_3\cdot p_j = \beta_{i,j}+\beta_{1}\cdot \overline{NrClaims}+\beta_{2}\cdot \overline{NutrVal}$ for all $i\in N$ and $j\in \{0,1\}$
\State Create a list $\mathcal{I}$ of indifference hyperplanes: $\beta_{i,0}-\beta_3\cdot p_0 = \beta_{i,1}-\beta_{3}\cdot p_1$ for all $i\in N$
\For{$h\in \mathcal{B}\cup\mathcal{I}$} 
    \For{$h'\in \mathcal{B}\cup\mathcal{I}$ such that $h'\neq h$} 
        \State Compute prices $p= h\cap h'$
        \State $\mathcal{P}:= \mathcal{P}\cup p$
    \EndFor
\EndFor    
\State {\bf Output}: $\mathcal{P}$
\end{algorithmic}
\end{algorithm}

Given a set of potentially optimal pricing strategies of the company, it is possible to compute the optimal sugar tax rate. Since the number of pricing strategies and, consequently, the number of consumer responses is finite, the social welfare function is a staircase function in $\alpha$ with break points possible only in the break-evens of the company's revenue:
$$\left(\sum_{i\in N}\ D_{i,0}\cdot p'_0\cdot x'_{i,0}\right)\ +\ (1-\alpha)\cdot \left(\sum_{i\in N}\ D_{i,1}\cdot p'_1\cdot x'_{i,1}\right) =$$ 
$$\left(\sum_{i\in N}\ D_{i,0}\cdot p''_0\cdot x''_{i,0}\right)\ +\ (1-\alpha)\cdot \left(\sum_{i\in N}\ D_{i,1}\cdot p''_1\cdot x''_{i,1}\right),$$    
where $p'$ and $p''$ are two potentially optimal pricing strategies, and $x'$ and $x''$ are the respective consumer choices. Evaluating the social welfare in every break point, and choosing for $\alpha$ the break point that maximizes the social welfare does solve the three-level mathematical program. We call this procedure Algorithm 2. Overall complexity of Algorithm 2 is $O(n^{2m})$ for $n$ consumers and $m$ products, which is still polynomial if the number of products $m$ is fixed. 
\begin{algorithm} [h!]
\scriptsize
\caption{Computing optimal sugar tax rate}\label{alg:optrate}
\begin{algorithmic}[1]
\State {\bf Input}: $\mathcal{P}$ from Algorithm 1, $D_{i,j}$ for all $i\in N,\ j\in \{0,1\}$ 
\State $\alpha^*:=0$ \Comment Optimal sugar tax rate
\For{$p\in\mathcal{P}$}
    \For{$i\in N$}
        \If {$u_{i,0}(p)\geq u_{i,1}(p)$} 
        \State $x_{i,0}(p)=1$ and $x_{i,1}(p)=0$ 
        \Else 
        \State $x_{i,1}(p)=1$ and $x_{i,0}(p)=0$
        \EndIf
        \If {$\max_{j\in\{0,1\}} u_{i,j}<0$} $x_{i,0}(p)=0$ and $x_{i,1}(p)=0$ \EndIf
    \EndFor
\EndFor
\State $W^*:=\max_{p\in \mathcal{P}} U_c(p,x(p))+U_f(p,x(p))$ \Comment Optimal social welfare
\For{$p'\in\mathcal{P}$}
    \For{$p''\in\mathcal{P}$}
        \State Compute $0\leq \alpha\leq 1$ such that 
        $$\left(\sum_{i\in N}\ D_{i,0}\cdot p'_0\cdot x_{i,0}(p')\right)\ +\ (1-\alpha)\cdot \left(\sum_{i\in N}\ D_{i,1}\cdot p'_1\cdot x_{i,1}(p')\right) =$$ 
        $$\left(\sum_{i\in N}\ D_{i,0}\cdot p''_0\cdot x_{i,0}(p'')\right)\ +\ (1-\alpha)\cdot \left(\sum_{i\in N}\ D_{i,1}\cdot p''_1\cdot x_{i,1}(p'')\right)$$
        \State $$p(\alpha):=\arg\max_{p\in \mathcal{P}} \sum_{i\in N}\ D_{i,0}\cdot p_0\cdot x_{i,0}(p)+(1-\alpha)\cdot \sum_{i\in N}\ D_{i,1}\cdot p_1\cdot x_{i,1}(p)$$
        \State $$W(\alpha):=\sum_{i\in N}\sum_{j\in M}\ \left(\ln (1+u_{i,j}(p(\alpha)))+D_{i,j}\cdot p_j(\alpha)\right)\cdot x_{i,j}(p(\alpha))$$
        \If{$W(\alpha)\geq W^*$}\ $\alpha^*=\alpha$ and $W^*:=W(\alpha)$ \EndIf
    \EndFor
\EndFor
\State {\bf Output}: $\alpha^*$ and $W^*$
\end{algorithmic}
\end{algorithm}

Notice, the approach remains polynomial even for the multi-level sugar taxes if the number of levels $L$ is also fixed. In this case, one has to determine not a single optimal $\alpha$, but an optimal set of $2L-1$ parameters: $L$ tax levels and $L-1$ sugar containment percentages where the tax jumps from one level to another. A break-even for the company's revenue is computed then similarly to the computation of the vertices of the choice regions in the price space, but now it is a vertex in the \emph{regulatory space} $R^{2L-1}_{+}$, where the revenues of $2L$ potentially optimal pricing strategies meet each other. The overall time complexity of the algorithm in this case is $O(n^{2mL})$, which is still polynomial in the input size. 

\section{Illustrative intuitive example }\label{example} 
The following example describes the process of obtaining an optimal sugar tax rate from the purchase data set. For simplicity of presentation, consider a market with two basic products, say Coca-Cola ($j=1$) and Coca-Cola Zero ($j=0$), where the first one contains sugar and the second one is sugar-free. For a generalization of this case one may consider two generic types of products and deal with the averages per product type. Given a purchase data set and applying multinomial choice model~\cite{healthclaims2017}, we obtain utility functions for three types of consumers, see Table~\ref{tab:utilities}.   
\begin{table}[h!]
\footnotesize
\centering
\begin{tabular}{l|l|l} 
& Coca-Cola & Coca-Cola Zero\\
\hline
(H)igh sugar & $u_{H,1}=(0.94-0.2 p_1)^+$ & $u_{H,0}=(0.41-0.26 p_0)^+$\\
consumers&&\\ 
(like sugar)&&\\
\hline
(M)edium sugar & $u_{M,1}=(0.17-0.18 p_1)^+$ & $u_{M,0}=(0.47-0.24 p_{0})^+$\\
consumers&&\\
(indifferent to sugar)&&\\
\hline
(L)ow sugar &  $u_{L,1}=(0.53-0.23 p_{1})^+$ & $u_{L,0}=(0.93-0.17 p_{0})^+$\\
consumers&&\\ 
(do not like sugar)&&
\end{tabular}\caption{\label{tab:utilities} Consumer preferences in soft drinks with different sugar content}
\end{table}

\begin{figure}[h!]
\center{\includegraphics[scale=0.9]{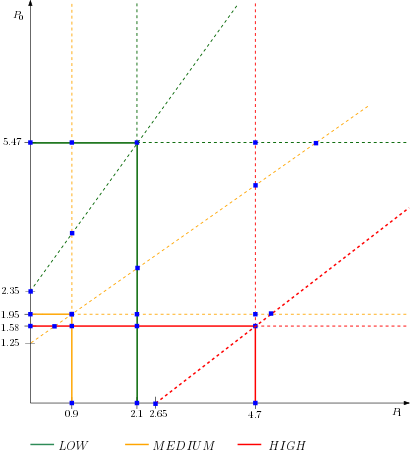}}
\caption{Price space}\label{fig:prices}
\end{figure}

When all budget and indifference lines/hyperplanes are drawn in the price space ($R^2$ in our case), we obtain the diagram as depicted in Figure~\ref{fig:prices}. The crossing points of the budget and indifference lines are the potential optimal pricing strategies of the company (Coca-Cola). Having three lines per consumer, we have nine lines in total, which might lead to at most 36 crossings, which will be the output of Algorithm 1. However, not all pairs of the lines cross each other, e.g., the budget lines for a product are parallel. This shrinks the number of potentially optimal pricing strategies to 27 points listed in Table~\ref{tab:points} together with consumer preferences (``CC'' stands for Coca-Cola and ``Z'' stands for Coca-Cola Zero), realized utilities of consumers and respective company revenues. In the case at hands, ``Low'' consumers have demand 11441 units, ``Medium'' consumers have demand 9433 units, and ``High'' consumers have demand 9942 units independent on the product (Regular or Zero). This allows us to compute the revenue of the company for every potentially optimal pricing strategy. As an intermediate result, we obtain that if the sugar tax rate $\alpha=0$, then the maximal revenue is 1093019.67 currency units (CU) in the following point: Coca-Cola is priced at 4.7 CU and Coca-Cola Zero is priced at 5.47 CU.

\begin{table}[h!]
\footnotesize
\centering
\begin{tabular}{l|l|l|l|l|l} 
Nr & Coordinates & High sugar & Medium sugar & Low sugar & Revenue \\\hline
1 & (0; 1.25) & $CC: u=0.94$ & $CC: u=0.17$ & $Z: u=0.72$ & $14301.25$ \\
2 &(0; 1.58) & $CC: u=0.94$ & $CC: u=0.17$ & $Z: u=0.66$ & $18076.78$ \\
3 & (0; 1.96) & $CC: u=0.94$ & $CC: u=0.17$ & $Z: u=0.6$ & $22424.36$ \\
4 & (0; 2.35) & $CC: u=0.94$ & $CC: u=0.17$ & $CC: u=0.53$ & $0$ \\
5 & (0; 5.47) & $CC: u=0.94$ & $CC: u=0.17$ & $CC: u=0.53$ & $0$ \\
6 & (0.44; 1.58) & $CC: u=0.85$ & $CC: u=0.09$ & $Z: u=0.66$ & $26601.78$ \\
7 & (0.94; 0) & $CC: u=0.75$ & $Z: u=0.47$ & $Z: u=0.93$ & $9345.48$ \\
8 & (0.94; 1.58) & $CC: u=0.75$ & $Z: u=0.0.9$ & $Z:0.66$ & $42326.4$ \\
9 & (0.94; 1.96) & $CC: u=0.75$ & $CC: u=0$ & $Z: 0.6$ & $40636.86$ \\
10 & (0.94; 3.62) & $CC: u=0.75$ & $CC: u=0$ & $CC: 0.31$ & $28967.04$ \\
11 & (0.94; 5.47) & $CC: u=0.75$ & $CC: u=0$ & $CC: u=0.31$ & $28967.04$ \\
12 & (2.13; 0) & $CC: u=0.51$ & $Z: u=0.47$ & $Z: u=0.93$ & $21176.46$ \\
13 & (2.13; 1.58) & $CC: u=0.51$ & $Z: u=0.09$ & $Z: u=0.66$ & $54157.38$ \\
14 & (2.13; 1.96) & $CC: u=0.51$ & $Z: u=0$ & $Z: u=0.6$ & $62089.5$ \\
15 & (2.13; 2.85) & $CC: u=0.51$ & $–$ & $Z: u=0.45$ & $53783.31$ \\
16 & (2.13; 5.47) & $CC: u=0.51$ & $–$ & $CC: u=0.04$ & $45545.79$ \\
17 & (2.65; 0) & $CC: u=0.41$ & $Z: u=0.47$ & $Z: u=0.93$ & $26346.3$ \\
18 & (4.7; 0) & $Z: u=0.41$ & $Z: u=0.47$ & $Z: u=0.93$ & $0$ \\
19 & (4.7; 1.58) & $CC: u=0$ & $Z: u=0.09$ & $Z: u=0.66$ & $79708.32$ \\
20 & (4.7; 1.96) & $CC: u=0$ & $Z: u=0$ & $Z: u=0.6$ & $87640.44$ \\
21 & (4.7; 4.78) & $CC: u=0$ & $–$ & $Z: u=0.12$ & $101415.38$ \\
22 & (4.7; 5.47) & $CC: u=0$ & $–$ & $Z: u=0$ & $\textbf{109309.67}$ \\
23 & (4.7; 8.7) & $CC: u=0$ & $–$ & $–$ & $46727.4$ \\
24 & (5.19; 1.96) & $–$ & $Z: u=0$ & $Z: u=0.6$ & $4913.04$ \\
25 & (5.63; 5.47) & $–$ & $–$ & $Z: u=0$ & $62582.27$ \\
26 & (9.75; 5.47) & $–$ & $–$ & $Z: u=0$ & $62582.27$ \\
27 & (16.31; 5.47) & $–$ & $–$ & $–$ & $–$
\end{tabular}
\caption{\label{tab:points} All potentially optimal points}
\end{table}

\newpage
Now, we are ready to calculate the optimal sugar tax rate $0\leq \alpha\leq 1$. We compute all break points of the social welfare as described in Algorithm 2. For instance, take a pair of pricing strategies, say, 9 and 13: pricing strategy 9 is defined by $p'_1=0.94$ and $p'_0=1.96$, and pricing strategy 13 is defined by $p''_1=2.13$ and $p''_0=1.58$. In point 9, ``High'' and ``Medium'' consumers purchase Coca-Cola while ``Low'' consumer buys Zero. In point 13, ``High'' consumer purchases Coca-Cola, and ``Medium'' joins ``Low'' in her preference to Zero. The break-even in revenue under these two pricing strategies is achieved with $\alpha$ being a solution to the equation:
{\footnotesize
\begin{displaymath} 
11441\cdot 1.96 +\ (1-\alpha)\cdot (9433+9942)\cdot 0.94 =(11441+9433)\cdot 1.58\ +\ (1-\alpha)\cdot 9942\cdot 2.13,
\end{displaymath}
}
where the left hand side addresses the revenue at point 9 and the right hand side addresses the revenue at point 13. The solution to the equation is $\alpha=4.62$, which is beyond the global upper bound of 1 for the tax rate. Thus, this break point can be disregarded. We perform the same calculations for all pairs of the potentially optimal pricing strategies and derive all potentially optimal tax rates $\alpha$ in interval $[0,1]$. 
We list all these potentially optimal tax rates in Table~\ref{tab:rates}.
\begin{table}[h!]
\footnotesize
\centering
\begin{tabular}{|c|c|c|c|c|c|c|} 
0.00 & 0.09 & 0.10 & 0.121 & 0.124 & 0.13 & 0.15 \\\hline	
0.15 & 0.16 & 0.18 & 0.21 & 0.23 & 0.31 & 0.32 \\\hline	
0.37 & 0.38 & 0.46 & 0.49 & 0.511 & 0.513 & 0.514 \\\hline	
0.52 & 0.53 & 0.54 & 0.60 & 0.61 & 0.62 & 0.63 \\\hline
0.69 & 0.7 & 0.61 & 0.83 & 0.84 & 0.97 & 1.00 
 \end{tabular}
\caption{\label{tab:rates} All potentially optimal tax rates $\alpha$}
\end{table}

For each potentially optimal tax rate, we calculate the company's maximal utility (revenue minus tax) by enumerating over all potentially optimal pricing strategies. In the case at hands, for any $0\leq \alpha\leq 1$, the maximal utilities of the company appear only in three price points: (22) Coca-Cola is priced at 4.7 CU, Coca-Cola Zero is priced at 5.47 CU; (25) Coca-Cola is priced 5.63 CU, Coca-Cola Zero is priced at 5.47 CU; (26) Coca-Cola is priced at 9.75 CU, Coca-Cola Zero is priced at 5.47 CU. Moreover, points 25 and 26 are optimal only in $\alpha=0$, while point 22 is optimal on the entire interval $[0;1]$.

Enumerating over all potentially optimal tax rates listed in Table~\ref{tab:rates}, we derive that the maximum social welfare of 156037 CU is achieved with $\alpha=1$, when the company sets the prices 4.7 CU for Coca-Cola and 5.47 CU for Coca-Cola Zero. 

\section{Conclusion}\label{sec:conclusion}
In this paper we develop a model for coordinating the interests of the government, companies and heterogeneous consumers. The model is based on a sequential game represented by a three-level mathematical program. We design an algorithm efficiently solving the program, i.e., obtaining a socially optimal solution in time polynomial in the input size of the problem. 

Surprisingly, for the case known in the literature~\cite{healthclaims2017}, we obtain that the optimal sugar tax rate maximizing the commonly used social welfare function is equal to 100\%. Furthermore, the real prices for Coca-Cola and Coca-Cola Zero are greatly underestimated compared to the prices that maximize the company (Coca-Cola) revenue. Moreover, the revenue maximizing prices with and without taxation are exactly the same. This phenomena might be caused by either oversimplification of the social welfare function, or by oversimplification of the consumer behavior model, or by inefficiency in the market caused by irrationality of the players, e.g., firms underpricing their products, or by a combination of the above factors. In this way, the approach proposed in the paper can be used in different contexts in order to benchmark the social welfare and also in order to check reasonability of the models/utilities of the players. 

\section*{Acknowledgment}\label{sec:acknowledgments}
We express our gratitude to Vladimir Kovalenok for the help in developing the example.

\end{document}